# Sleep Deprivation Attack Detection in Wireless Sensor Network

Tapalina Bhattasali
Techno India College of Technology
Kolkata, India
tapolinab@gmail.com

Rituparna Chaki
West Bengal University of Technology
Kolkata, India
rituchaki@gmail.com

Sugata Sanyal
Tata Institute of Fundamental Research
Mumbai, India
sanyal@tifr.res.in

## ABSTRACT
Deployment of sensor network in hostile environment makes it mainly vulnerable to battery drainage attacks because it is impossible to recharge or replace the battery power of sensor nodes. Among different types of security threats, low power sensor nodes are immensely affected by the attacks which cause random drainage of the energy level of sensors, leading to death of the nodes. The most dangerous type of attack in this category is sleep deprivation, where target of the intruder is to maximize the power consumption of sensor nodes, so that their lifetime is minimized. Most of the existing works on sleep deprivation attack detection involve a lot of overhead, leading to poor throughput. The need of the day is to design a model for detecting intrusions accurately in an energy efficient manner. This paper proposes a hierarchical framework based on distributed collaborative mechanism for detecting sleep deprivation torture in wireless sensor network efficiently. Proposed model uses anomaly detection technique in two steps to reduce the probability of false intrusion.

## 1. INTRODUCTION
Most devastating sleep deprivation torture comes in the form of sending useless control traffic and forces the nodes to forgo their sleep cycles so that they are completely exhausted and hence stop working. This type of attack is difficult to detect because of its apparently innocent nature. In resource constrained wireless sensor network, traditional security mechanisms fail to provide the necessary protection for sensed data. The absence of infrastructure also makes it difficult to detect security threats. Therefore security mechanisms have to be designed with efficient resource utilization, especially power. In wireless sensor network, maximum security can only be achieved by designing an effective detection model whose purpose is to provide alert about possible attacks, ideally in time to stop the attack or to mitigate the damage. It attempts to differentiate abnormal activities from normal ones, and identify malicious activities. Generally anomaly based detection mechanism has the intelligence to detect variations from normal behaviors and respond to new intrusions, whereas pattern based detection mechanisms have the capability to identify all known intrusions accurately.

Intrusion detection for wireless sensor network is an emerging field of research. In one of our earlier works, we have presented a survey of recent IDS in sensor networks [1]. In stand-alone IDS, each node independently determines intrusions. It gives rise to problem in handling dense network. Therefore stand-alone IDS is not opted for research work in sensor network. Distributed ID mechanism [2,3] can be feasibly implemented in sensor network because load of detection is distributed which reduces single node overhead, and gives better performance in terms of the important parameters such as energy consumption, response time, detection accuracy. But no efficient IDS is found to detect sleep deprivation torture accurately. For this reason, here we propose a distributed collaborative detection model based on layered architecture.

The remainder of this paper is organized as follows. Section 2 explains the relevant work in this field. Section 3 consists of an outline of the proposed system model. Section 4 discusses the case study of the proposed model. Section 5 illustrates the comparative analysis of the proposed model. It is followed by a conclusion in section 6. Section 7 gives reference list of this paper.

## 2. RELEVANT WORK
In this paper, we focuses on sleep deprivation attack which is also considered as layer 2 attack. This section gives an idea about the related mitigation technique of it.

The network lifetimes of existing Medium Access Control (MAC) protocols such as Sensor MAC (S-MAC), Timeout MAC (T-MAC) and Berkley MAC (B-MAC) were compared by Raymond et. al.[4]. Brownfield et. al. [5] had proposed a protocol Gateway MAC to mitigate the effects of denial of sleep attacks. WSNET link layer protocol G-MAC can serve as an effective denial of sleep defense by centralizing cluster management.

 In [6], it is assumed that adversary nodes must become cluster heads in order to launch sleep deprivation attack. Three separate methods are analyzed for mitigating sleep deprivation attack: the random vote scheme, the round robin scheme, and the hash-based scheme. Pirretti et. al. have evaluated these schemes based upon their ability to reduce the adversary's attack, the amount of time required to select a cluster head and the amount of energy required to perform each scheme. It have been found that, hash-based scheme is the best among three clustering methods at mitigating the sleep deprivation attack in terms of resilience towards attack and required overhead.

In [7], a host based lightweight intrusion detection technique, Clustered Adaptive Rate Limiting (CARL) based on rate limiting approach at MAC layer is proposed to defeat denial-of-sleep attacks. In this adaptive rate limiting approach, network traffic is restricted only when sufficient malicious packets have been sensed to suspect that the network is under attack. It can be used to maintain network lifetimes and better throughput at a time even in the face of sleep deprivation attack.

In [8], a scheme is proposed employing fake schedule switch with RSSI measurement aid. The sensor nodes can reduce and weaken the harm from exhaustion attack and on the contrary make the attackers lose their energy quickly so as to die.

In [9], a quickest intrusion detection scheme, modeled as Markov Decision Process (MDP) has been proposed by keeping a minimal number of sensors active. Three sleep/wake scheduling algorithms are mentioned here.
i) optimal control of the number of sensors in the wake state in a time slot.
ii) optimal control of the probability of a sensor in the wake state in a time slot.
iii) optimal probability of a sensor in the wake state.
It ensures that energy expenditure for sensing, computation and communication is minimized and the lifetime of the network is maximized.

## 3. PROPOSED MODEL

Most of the existing detection techniques have not met the requirements for practical deployment in wireless sensor network to mitigate sleep deprivation torture. In this section, a hierarchical model is proposed for wireless sensor network to detect the sensor nodes affected by sleep deprivation attack. It uses cluster based mechanism in an energy efficient manner. A dynamic detection model is designed here to overcome sudden death of IDS enabled sensor nodes. In this model responsibility of each node dynamically changes to reduce the burden of a single node.

Our research focuses on distributed anomaly detection technique in order to provide a reliable and energy efficient heterogeneous wireless sensor network. Anomaly is detected by comparing the values with predefined parameters specified in normal profile. The proposed model uses anomaly detection technique in such a way so that false intrusion detection can be avoided. To mitigate the attack, proposed model physically excludes malicious nodes from the network and rejects fake packets.

In heterogeneous sensor field, sensor nodes are categorized into various roles such as sink gateway (SG), cluster-in-charge (CIC), sector monitor(SM), sector-in-charge (SIC) and leaf node (LN) depending on their battery capacity. The roles of CIC, SM and SIC are changed dynamically to avoid exhaustion of nodes. Sink Gateway node is the honest gateway to another network or access point. SG is preset to perform gateway functionality.

According to the working principle of the proposed model, the heterogeneous sensor network is partitioned into clusters which are again partitioned into sectors to conserve energy and avoid redundant exchange of messages among the sensor nodes. This helps in prolonging the battery life of the individual sensor node, so as to reduce the rate of energy consumption and increase network lifetime. Detection model can be loaded at each node, but if the nodes are designated as sector monitor, only anomaly detector module is activated. Nodes designated as cluster-in-charge have the rights to activate decision maker module in addition of anomaly detector module. Sensed data generated by LNs are transmitted to the access point through the sink gateway.

**Table 1. Participating Nodes in Wireless Sensor Network**

| Nodes | Definition |
|---|---|
| Sink Gateway(SG) | A layer 4 node having highest capacity provides gateway functionality to other networks. |
| Cluster-In-Charge(CIC) | A layer 3 node having maximum energy and degree (number of nodes within its coverage area) among all neighbors of SG and capable to take final decision regarding intrusion. |
| Sector Monitor(SM) | A layer 2 node that is nearest neighbor of CIC and whose detection power is set to maximum within sector and capable to detect anomaly. |
| Sector-In-Charge(SIC) | A layer 2 node that has maximum energy among neighbors of CIC and capable to collect sensing data. |
| Leaf Nodes (LN) | A layer 1 node which can only sense data and whose detection power is set to null. |

Table 1 gives the definition of participating nodes.

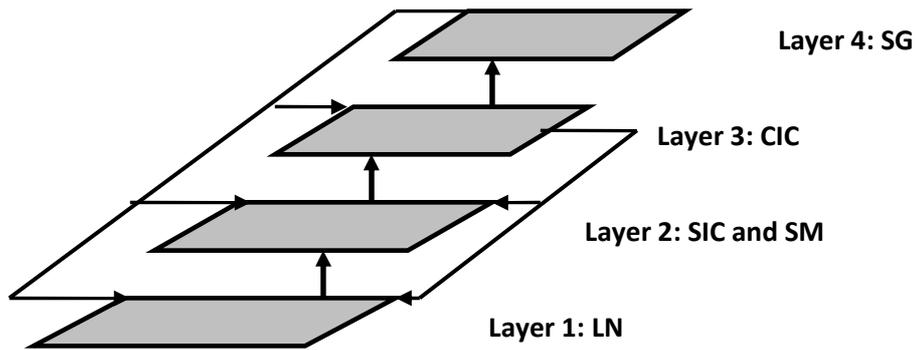

**Fig 1: Layers of Proposed Model**

Figure 1 represents the layers of the proposed model.

Sink Gateway (SG) broadcasts query message along with its own node-id to all the nodes within its range. According to the responses received from the neighboring nodes, SG prepares status profile of each of the neighbor nodes and sends to corresponding nodes. According to the capacity of nodes, Cluster-In-Charge (CIC) is selected and its coverage area is considered as cluster. CIC initiates data collection request to the node known as Sector-In-Charge (SIC) (whose coverage area is known as sector). SIC forwards data request to Leaf Node (LN) who returns sensing data to SIC. After collecting data from leaf nodes SIC forwards data to Sector Monitor(SM). SM checks whether data is valid or suspected and marks data packets accordingly before forwarding to CIC. Final decision is taken by CIC. Genuinely valid data is forwarded to SG and corrupted data is rejected. In the middle of the working period if any new node is discovered by SG, it sends sleep signal to the node to avoid unnecessary wastage of node's battery.

Data traffic is destined for SG, so every data packet at a node is forwarded to the node's parent in the tree rooted at the SG. According to the data collection model, sensing data are collected by SIC from LNs. SIC forwards data to SM who checks whether data are anomalous or not. If anomaly is found in the data and sending node's behavior is found to be suspected, then INVALID tag is set for the packet, otherwise VALID tag is set. Anomalous packet is detected if number of packets within a specific time interval exceeds threshold limit or packet is transmitted within sleep schedule of the sender. Suspected node is detected by checking the residual energy of the sender with respect to normal residual energy. This is the first step of detection. SM sends all packets to CIC.INVALID packets and suspected sender nodes are further analyzed by CIC to take final decision of intrusion. This is the second step of detection to reduce the probability of false positive detection. CIC either forwards genuine data to SG or rejects fake packets. SG includes suspected node's id in its isolation list.

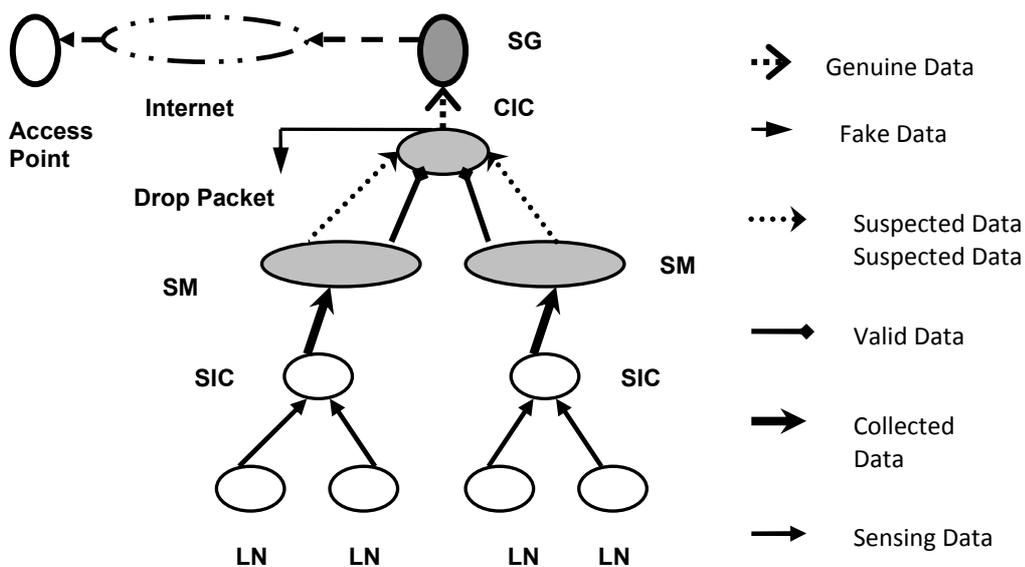

**Fig 2: Data Collection Model**

Figure 2 gives the idea about data collection procedure from source to destination in the proposed model. Figure 3 gives the idea about the overall work-flow of the proposed model.

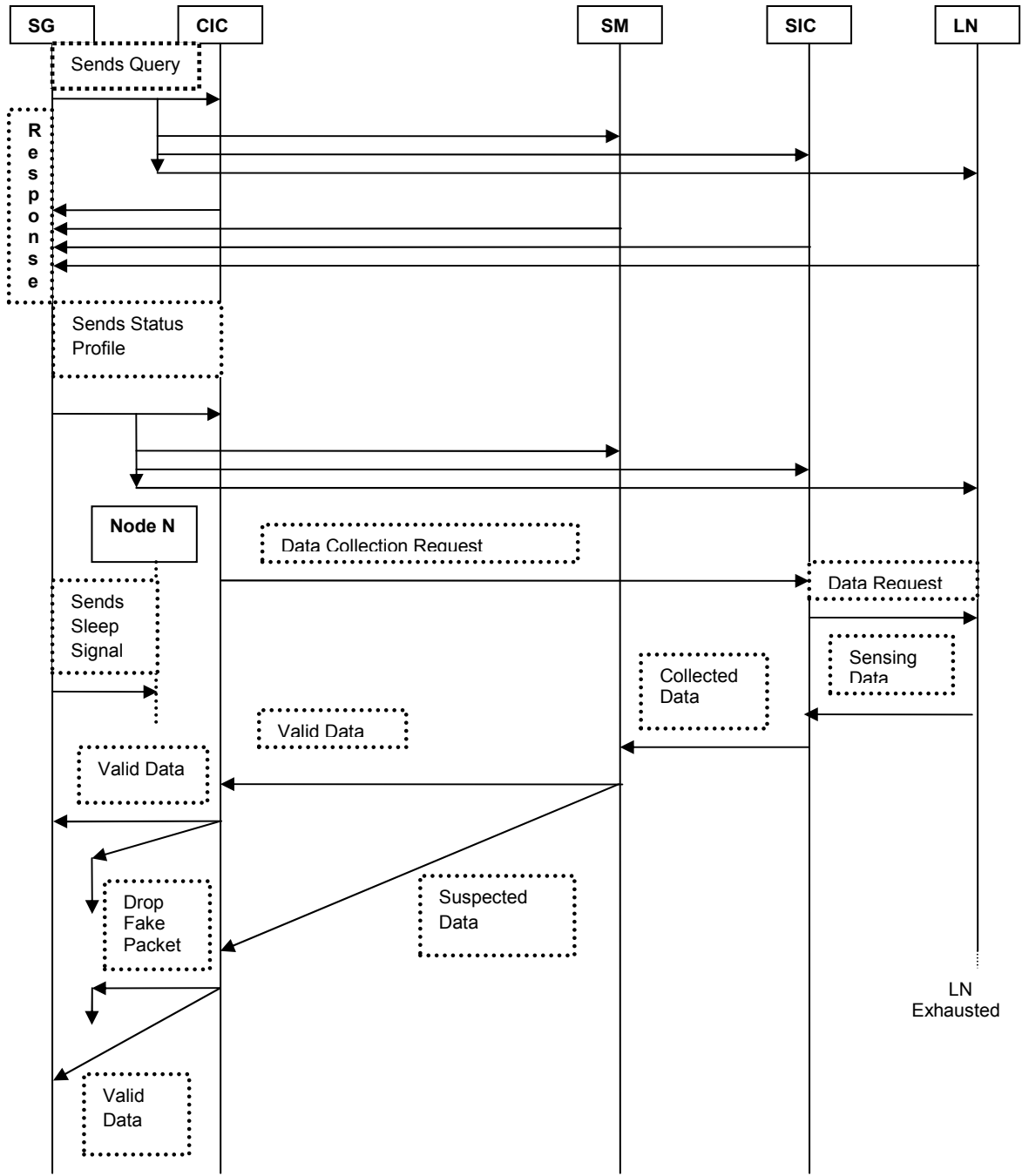

**Fig 3:    Procedural Logic of Proposed Model**

## 4. CASE STUDY

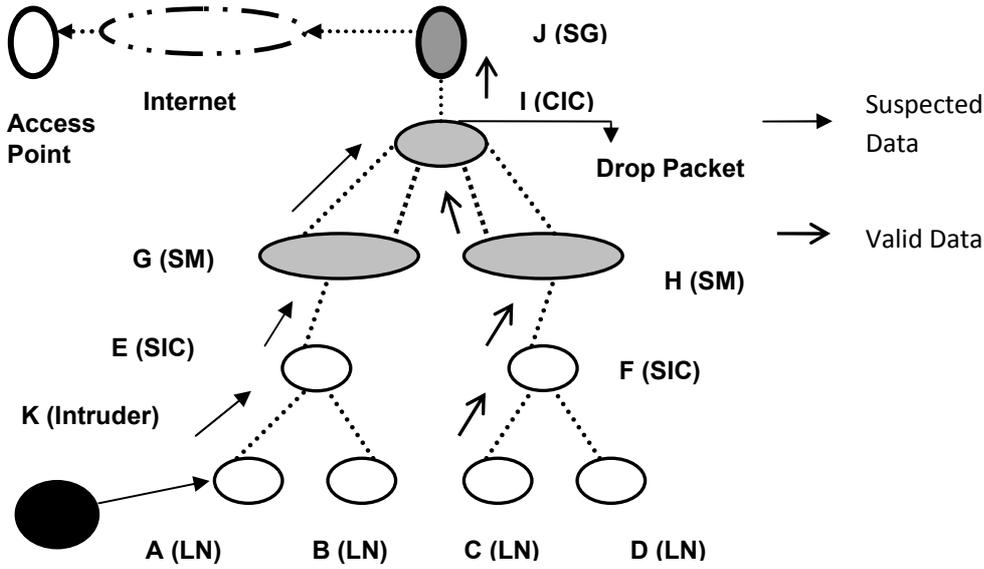

**Fig 4 :  A scenario where wireless sensor network is affected by sleep deprivation torture**

Figure 4 represents a case study of wireless sensor network according to our proposed model. It consists of number of sensor nodes such as A,B,C,D,E,F,G,H,I,J. Here node A, node B, node C and node D are designated as Leaf Node. Node E and node F are designated as Sector-In-Charge. Node G and node H are designated as Sector Monitor. Node I is designated as Cluster-In-Charge and node J is designated as Sink Gateway. Two cases are considered here.

**Case 1:** Assume the case where one unknown node K sends fake requests to Leaf Node A to force it to wake-up even in its sleep schedule. As a result node A sends packet P1 to SIC node E in its sleep schedule. SIC forwards packet P1 to SM node G. SM detects that packet P1 comes from sender node A within its sleep schedule $T_{slp}(A)$.SM detects packet P1 as anomalous and set status(P1) as INVALID. Then SM monitors the residual energy of node A.

As residual energy of node A is found to be less than threshold residual energy of the node, i.e.  Residual_Eng (A) < Th_RE, node A is considered as SUSPECTED.Packet P1 and related data are sent to CIC node I who analyzes packet again and takes final decision of intrusion.

Packet P1 is dropped and node A is isolated by including its id in isolation list maintained by SG.

**Case 2:** Assume another case where Leaf Node C sends packet P2 to SIC node F. Node F forwards packet P2 to SM node H. Node H detects no anomaly to packet P2 and set status(P2) as VALID. Packet P2 is forwarded to CIC node I who analyzes it as genuine and forwards the packet to SG node J to transmit the packet to its destination.

In the proposed model, it is assumed that leaf nodes are directly affected by intruder. If any leaf node receives fake data request from unknown nodes (intruder), it can not detect it. As a result battery of the affected node may be low or exhausted completely. This can affect data transmission of the network. For this reason, data transmission must be done in authenticated way.

# 5. COMPARATIVE ANALYSIS OF SLEEP DEPRIVATION ATTACK DETECTION TECHNIQUES

**Table 2. Strengths and Weakness of Detection Techniques**

| Detection Techniques | Strengths | Weakness |
|---|---|---|
| Sensor MAC (S-MAC) [4] | Simple energy-efficient protocol extends WSN network lifetime. | (i) Fixed sleep cycle makes it vulnerable to broadcast as well as unicast attack. (ii) Inflexible in responding to network traffic fluctuations or network scaling. |
| Timeout MAC (T-MAC) [4] | (i) Dynamic sleep cycle makes network flexible and scalable. (ii) Energy saving is comparatively better. | It is more vulnerable to a broadcast attack. |
| Berkley MAC (B-MAC) [4] | It works well in ultra-low traffic networks. | Performance significantly decreases because each passive node has to wake up and receive every message. |
| Gateway MAC (G-MAC) [5] | It performs significantly better than the other in every traffic situation. | All cluster nodes entirely dependent on gateway node. |
| Random Vote Scheme [6] | It reduces probability of selecting adversary cluster head; so that exhaustion of sensor nodes by cluster head is reduced. | (i) It requires more iteration to complete the algorithm. (ii) When number of compromised nodes within a cluster increases, attack detection becomes difficult. (iii) It is less scalable. |
| Round Robin Scheme [6] | (i) In this, adversary node can not easily declare itself as cluster head. (ii) It requires single iteration to select cluster head. | For large cluster, each node requires an unrealistic amount of per-node storage, which enhances the overhead. |
| Hash based Scheme [6] | (i) It is highly scalable. (ii) Energy required to select cluster head is minimum. (iii) It is very efficient to mitigate attack in terms of overhead. | It only considers intrusion from cluster head side. |
| Clustered Adaptive Rate Limiting (CARL) [7] | (i) Lightweight detection mechanism makes it efficient in resource limited network. (ii) Network lifetime can be maintained even in the face of attack. | High throughput may not be maintained at all time because of the limitation in radio activation duration. |
| RSSI Measurement Aid [8] | (i) Fake schedule switch scheme successfully defends attack by exhausting energy of attacker quickly. (ii) Network health is maintained. | Energy consumption and transmission delay increase. |
| Markov Decision Process (MDP) [9] | (i) To increase network lifetime, energy expenditure becomes low by keeping a minimal number of sensors active. (ii) Attack detection becomes quick with a minimal number of observations by maintaining low cost. | Procedural complexity is high. |
| Proposed Model | (i) It enhances energy efficiency and network scalability by using | Packet transmission overhead may become high in some cases. |

| | clusterization and sectorization approach. (ii)Distributed collaborative mechanism increases network lifetime. (iii)Two level detection mechanism can efficiently detect attack by reducing probability of false detection. | |
|---|---|---|

## 6. CONCLUSION

In this paper, an effort has been made to propose a collaborative hierarchical model capable of detecting insomnia of sensor nodes [1,10]. The aim of proposed model is to save the power consumption of sensor nodes so as to extend the lifetime of the network, even in the face of sleep deprivation torture. Proposed model virtually eliminates the probability of phantom detection [11] by using two phase detection procedure. Workload of the proposed model is distributed among the components according to their capacity to avoid complete exhaustion of battery power. Analysis of the proposed model shows that it is efficient in detecting sleep deprivation attack in wireless sensor network.